\documentclass{elsart}
\input epsf
\usepackage{latexsym}
\begin{document}
\begin{frontmatter}
\title{Strange Meson Enhancement in PbPb Collisions }
\collab{The NA44 Collaboration}
\author[NBI]{I. Bearden}, 
\author[NBI]{H. B{\o}ggild}, 
\author[LANL]{J. Boissevain},
\author[Nant]{L. Conin}, 
\author[Colu]{J. Dodd},
\author[Nant]{B. Erazmus}, 
\author[Hiro]{S. Esumi \thanksref{Heidelberg}}, 
\author[CERN]{C. W. Fabjan}, 
\author[Zagr]{D. Ferenc}, 
\author[LANL]{D. E. Fields \thanksref{UNM}},
\author[CERN]{A. Franz \thanksref{Brookhaven}}, 
\author[NBI]{J. J. Gaardh{\o}je}, 
\author[NBI]{A. G. Hansen}, 
\author[NBI]{O. Hansen}, 
\author[OSU]{D. Hardtke \thanksref{LBNL}}, 
\author[LANL]{H. van Hecke}, 
\author[CERN]{E. B. Holzer}, 
\author[OSU]{T. J. Humanic}, 
\author[CERN]{P. Hummel}, 
\author[SUNY]{B. V. Jacak}, 
\author[OSU]{R. Jayanti},
\author[Hiro]{K. Kaimi \thanksref{deceased}}, 
\author[Hiro]{M. Kaneta}, 
\author[Hiro]{T. Kohama},
\author[SUNY]{M. L. Kopytine \thanksref{FIAN}}, 
\author[Colu]{M. Leltchouk}, 
\author[Zagr]{A. Ljubicic, Jr.}, 
\author[Lund]{B. L{\"o}rstad}, 
\author[Hiro]{N. Maeda \thanksref{FSU}}, 
\author[Nant]{L. Martin},
\author[Colu]{A. Medvedev}, 
\author[Texa]{M. Murray}, 
\author[Hiro]{H. Ohnishi \thanksref{Brookhaven}}, 
\author[CERN]{G. Paic}, 
\author[OSU]{S. U. Pandey \thanksref{Wayne}},
\author[CERN]{F. Piuz}, 
\author[Nant]{J. Pluta \thanksref{Warsaw}}, 
\author[BNL]{V. Polychronakos}, 
\author[Colu]{M. Potekhin}, 
\author[CERN]{G. Poulard}, 
\author[OSU]{D. Reichhold}, 
\author[Hiro]{A. Sakaguchi \thanksref{Osaka}}, 
\author[Lund]{J. Schmidt-S{\o}rensen}, 
\author[LANL]{J. Simon-Gillo},
\author[LANL]{W. Sondheim}, 
\author[Hiro]{T. Sugitate},
\author[LANL]{J. P. Sullivan}, 
\author[Hiro]{Y. Sumi}, 
\author[Colu]{W. J. Willis}, 
\author[Texa]{K. L. Wolf \thanksref{deceased}}, 
\author[LANL]{N. Xu \thanksref{LBNL}},
\author[OSU]{D. S. Zachary} 
\address[NBI]{Niels Bohr Institute, DK-2100, Copenhagen, Denmark}
\address[LANL]{Los Alamos National Laboratory, Los Alamos, NM 87545,
 USA}
\address[Colu]{Columbia University, New York, NY 10027, USA}
\address[Nant]{Nuclear Physics Laboratory of Nantes, 44072 Nantes,
 France}
\address[Hiro]{Hiroshima University, Higashi-Hiroshima 739, Japan}
\address[CERN]{CERN, CH-1211 Geneva 23, Switzerland}
\address[Zagr]{Rudjer Boscovic Institute, Zagreb, Croatia}
\address[Texa]{Texas A\&M University, College Station, Texas 77843,
 USA}
\address[OSU]{Ohio State University, Columbus, OH 43210, USA}
\address[SUNY]{SUNY at Stony Brook, Stony Brook, NY 11794, USA}
\address[Lund]{University of Lund, S-22362 Lund, Sweden}
\address[BNL]{Brookhaven National Laboratory, Upton, NY 11973, USA}
\thanks[deceased]{deceased}
\thanks[Heidelberg]{now at Heidelberg University, D-69120 Heidelberg,
 Germany}
\thanks[UNM]{now at University of New Mexico, Albuquerque, NM 87131,
 USA}
\thanks[Brookhaven]{now at Brookhaven National Laboratory, Upton,
 NY 11973, USA}
\thanks[LBNL]{now at Lawrence Berkeley National Laboratory, 
              Berkeley, CA 94720, USA}
\thanks[FIAN]{on an unpaid leave from P. N. Lebedev Physical Institute,
Russian Academy of Sciences}
\thanks[FSU]{now at Florida State University, Tallahassee, FL 32306,
 USA}
\thanks[Wayne]{now at Wayne State University, Detroit MI 48201, USA}
\thanks[Warsaw]{Institute of Physics, Warsaw University of Technology,
  Koszykowa 75, 00-662 Warsaw, Poland}
\thanks[Osaka]{now at Osaka University, Toyonaka, Osaka 560-0043, 
Japan}

\begin{abstract} 
The NA44 Collaboration has measured yields and
differential distributions of $K^+$, $K^-$, $\pi^+$, $\pi^-$ in
transverse kinetic energy and rapidity, around the center-of-mass
rapidity in 158 A GeV/c Pb$+$Pb collisions at the CERN SPS.  A
considerable enhancement of $K^+$ production per $\pi$ is observed, as
compared to $p+p$ collisions at this energy.  To illustrate the
importance of secondary hadron rescattering as an enhancement
mechanism, we compare strangeness production at the SPS and AGS with
predictions of the transport model RQMD.  \end{abstract}
\begin{keyword} 
Relativistic heavy ion collisions; Particle
production; Kaon; Pion; Strangeness enhancement; Quark-Gluon Plasma.
\end{keyword} 

\end{frontmatter}

\section{Introduction}

 Ultrarelativistic heavy ion collisions create a
highly excited complex system, whose dynamics are governed by
excitation of nucleonic, mesonic, resonance \cite{RQMDf} and, to some
unknown extent, quark and gluon degrees of freedom.  It has been
predicted that the extreme conditions of temperature and density in
such collisions may suffice to create a state, known as quark-gluon
plasma (QGP), where the quarks are no longer confined in hadrons
\cite{QGP}.  This has stimulated experimental searches for evidence of
the deconfinement phase transition.  Interactions between liberated
gluons in the deconfined phase are predicted \cite{RafHag} to enhance
the rate of strangeness production compared to the non-QGP scenarios.

Being the lightest strange hadrons, kaons are expected to dominate
the strange sector by virtue of canonical 
thermodynamics \cite{bel}.
The observed kaon multiplicity yields information about the mechanism 
of strangeness production, hadronization and 
subsequent evolution in the hadron gas, before
the gas becomes sufficiently dilute that the interactions cease.
Inelastic hadronic rescattering can enrich the strangeness content
of the system \cite{koch}.
We report the yields and distributions of charged  kaons  and pions 
measured in ultrarelativistic PbPb collisions by the NA44 Experiment,
and discuss implications of these data on the physics of the 
above-mentioned hadronic processes.

\section{Experiment and data analysis}

The NA44 Collaboration has measured PbPb collisions at
158 A GeV/c using a focusing spectrometer at the CERN SPS.
A magnet system of two dipoles and three focusing
quadrupoles, together with a tracking complex (a pad chamber,
three highly segmented scintillation hodoscopes H2, H3, H4 and two
strip chambers) provides momentum resolution of 0.2\%.
The spectrometer accepts charged particles of a single charge at a 
time,
has two angular positions (44 and 131 mrad) and is
operated at two different field strengths.
In the weak field mode, it accepts charged tracks in the
momentum range of $3.3<p<5.1$ GeV$/c$, 
and of $6.3<p<9.7$ GeV$/c$ in the strong field mode.
These two field modes are often called ``the 4 GeV/c'' and 
``the 8 GeV/c'' settings, respectively.
More details about the spectrometer are given in \cite{NA44ex}.

Low (predominantly single track) multiplicity in the spectrometer
acceptance allows use of two Cherenkov counters (C1, C2) for threshold
discrimination of particles of different mass.
Collection of $K/p$  and $\pi$ samples uses
different trigger requirements: in the $K/p$ mode, the absence of
pions and electrons
in the acceptance is enforced by a Cherenkov veto (on C1, or both C1 
and C2, depending on the momentum setting), whilst for pions, 
no special trigger enrichment is needed.

Separation of kaons from protons in all settings is performed off-line
using the time-of-flight difference between $K$ and $p$.
The time-of-flight is measured using the beam counter 
\cite{bc} (with 35 ps resolution) as start and H3 
(with 100 ps resolution) as stop.
In the weak magnetic field mode,
the pions used in this analysis are identified by time-of-flight,
while
events with electrons in  the acceptance are rejected off-line using
C2.
High ($\ge 98 \%$) purity of the $K$ and $\pi$ samples is achieved.
In the strong field mode, pions are obtained by subtracting identified
kaons and protons from all charged tracks.

Inefficiencies due to the Cherenkov vetoes are evaluated 
by measuring the rejection by the Cherenkovs in untriggered runs.
Such unwanted vetoes occur when a kaon or proton is accompanied by
a pion, electron or muon in the Cherenkov counters.
To evaluate the inefficiency in the weak magnetic field runs,
the vetoed kaons are identified by time-of-flight
and the Uranium calorimeter \cite{UCAL} data is used for $\pi/e$ 
separation.
In the strong magnetic field runs, 
the momentum of the particles is too high for reliable separation
by time-of-flight, and
subtraction of pions, utilizing knowledge of
the pion line shape in $m^2$, is used to count vetoed kaons.

NA44 has two detectors to characterize event
multiplicity: $T_0$ (a scintillator trigger counter
 covering $ 1.4 \le \eta \le 3.7 $  for an $\eta$-dependent
fraction of azimuthal angle,
$0.22 \le \Delta \phi / 2\pi \le 0.84 $
respectively), and a Si pad array measuring $\,dE/\,dx$ in
512 pads covering  $1.5\le \eta \le 3.3 $ and $2\pi$ azimuthally.
The multiplicity of a given particle, measured in the spectrometer,
is an average over many events of a certain centrality class,
set by the trigger.
Accurate determination of the trigger centrality is performed
by varying the centrality used in normalizing 
the yield of charged tracks in the spectrometer 
until this yield agrees with the multiplicity in the Si array.
Correction for the acceptance difference between the spectrometer
 and the
Si array is performed using the RQMD model \cite{RQMD}, which is
consistent with measured \cite{NA49h-} charged hadron distributions.
Kaon and pion samples of identical Si multiplicity 
are selected via the $T_0$ signal amplitude.

Differential distributions of particles in rapidity, $y$, and 
transverse
kinetic energy, $m_T-m$, carry information about the dynamics of the
collision. 
In determining $dN/dy$ for kaons and pions we use spectrometer 
settings, or portions thereof, with
$\Delta y = 0.2-0.6$.
Any dependence of the slope parameter(s) upon $y$ is therefore
negligible. Then 
\begin{equation}
{\Big\langle E\frac{\,d^3N}{\,dp^3}\Big\rangle}_{\Delta y, 2\pi} =
\frac{1}{2\pi}{\Big\langle\frac{\,dN}{m_T\,dm_{T}}
\Big\rangle}_{\Delta y} =
\frac{\frac{\,dn}{m_T\,dm_{T}}\int\limits_{\Delta y}^{} 
\frac{\,d\widetilde{N}}{\,dy} \,dy}
     {2\pi\Delta y \int\limits_{\Delta y}^{}A
     \frac{\,d\widetilde{N}}{\,dy} \,dy}
\end{equation}
where $A=A(y,m_{T})$ is the acceptance function
from Monte Carlo simulation of the spectrometer, including effects 
of magnetic
optics, detector response, momentum resolution, tracking efficiency 
and decays.
$\,dn/\,dm_T$ is the number density of reconstructed tracks in  $m_T$.
$d\widetilde{N}/dy$ is the shape of the rapidity distribution,
 taken to be Gaussian around midrapidity. 
  Integration of the $m_T$ distribution with extrapolation
  to $m<m_{T}<\infty$, using the fitted slopes,  results  in 
  $\,dN/\,dy$.  

Table \ref{sigma_rep} shows the sources of
uncertainty on $dN/dy$. 
The error in the extrapolation due to uncertainty in the slope 
parameter(s)
is small because over 95\% of particles around mid-rapidity have $p_T$
 in
the range covered by one of the two angle settings.
Consequently, the systematic error in $dN/dy$ is dominated not by the
extrapolation, but by uncertainties in determination of centrality
and particle ID efficiency. 
\begin{table}
\caption{Summary of fractional
systematic errors to the normalized yields. 
As the ``representative'' case, the case of positive kaons in weak
field
high angle spectrometer setting is chosen.
The data in the lines marked ``worst'' and ``best'' are not linked 
by the choice of a specific setting, but list the maximum and minimum 
uncertainty 
(among all settings)  due to a particular source.
\label{sigma_rep}}
\begin{tabular}{ccccc}
\hline
 case      & centrality & PID     & $p_T$ extrapolation & total \\
\hline

representative & 0.081  &  0.042  &          0.0067     & 0.098  \\
worst       &  0.081    &  0.25   &          0.063      & 0.26  \\
best        &  0.058    &  0.014  &          0.0042     & 0.058 \\
\hline
\end{tabular}
\end{table}

\section{Results and discussion}
\begin{figure}
\epsfxsize=7.5cm  
\epsfbox{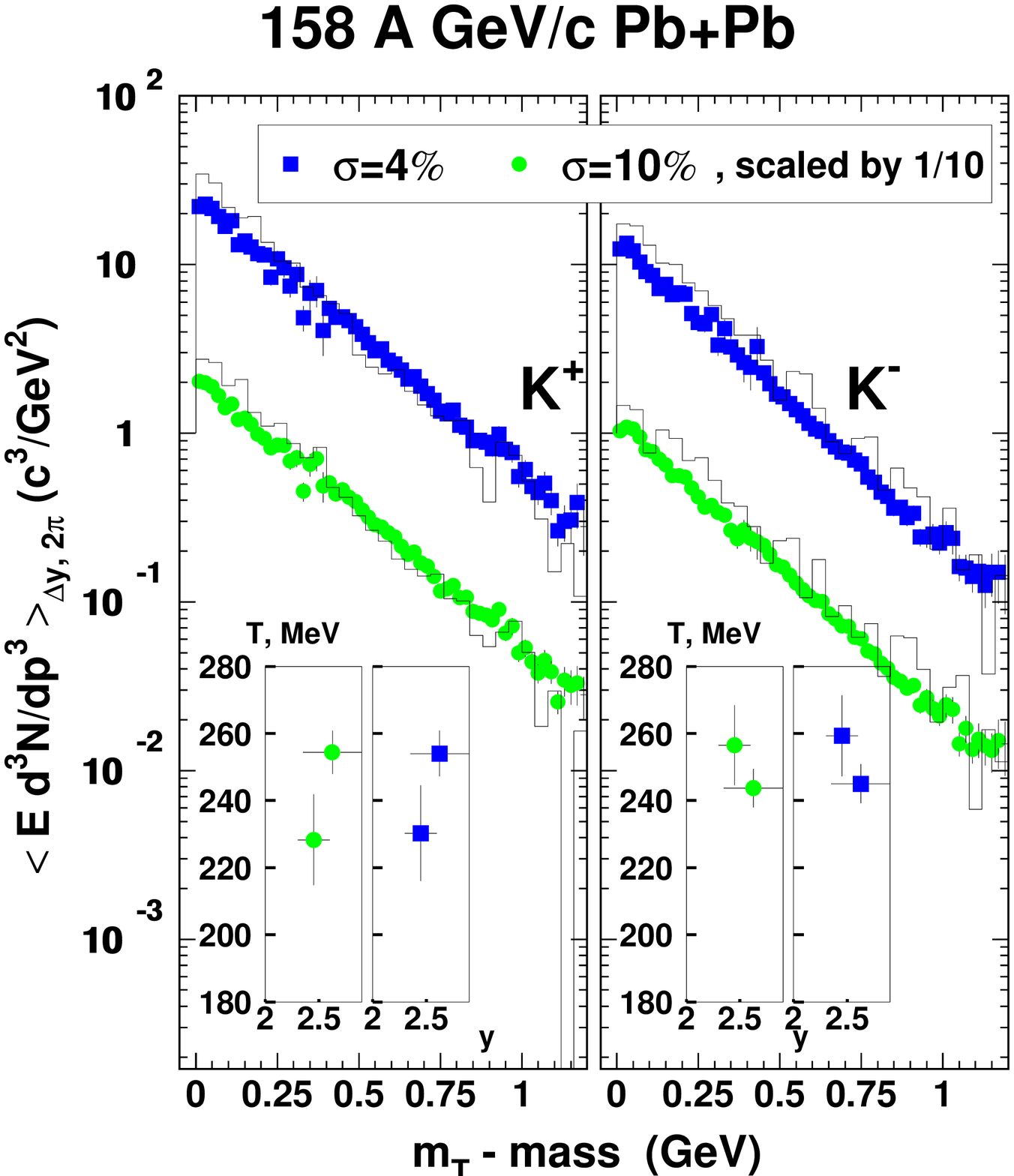}
\caption{Measured transverse kinetic energy distributions of 
positive and negative kaons
for the  4\% and  10\% most central of Pb+Pb collisions.
RQMD predictions for $|y-y_{CM}|<0.6$ (i.e., within NA44 acceptance)
are shown as  histograms. The fits follow the form
$1/m_{T}\,dN/\,dm_{T} \propto exp(-m_{T}/T)$, 
where $m_{T} = (m^{2}+p_{T}^{2})^{1/2}$.
$y$ ranges of the fits are given in Table 2 and are indicated by 
the horizontal
errorbars in the inserts.} 
\label{kt}
\end{figure}

\begin{figure}
\epsfxsize=7.5cm 
\epsfbox{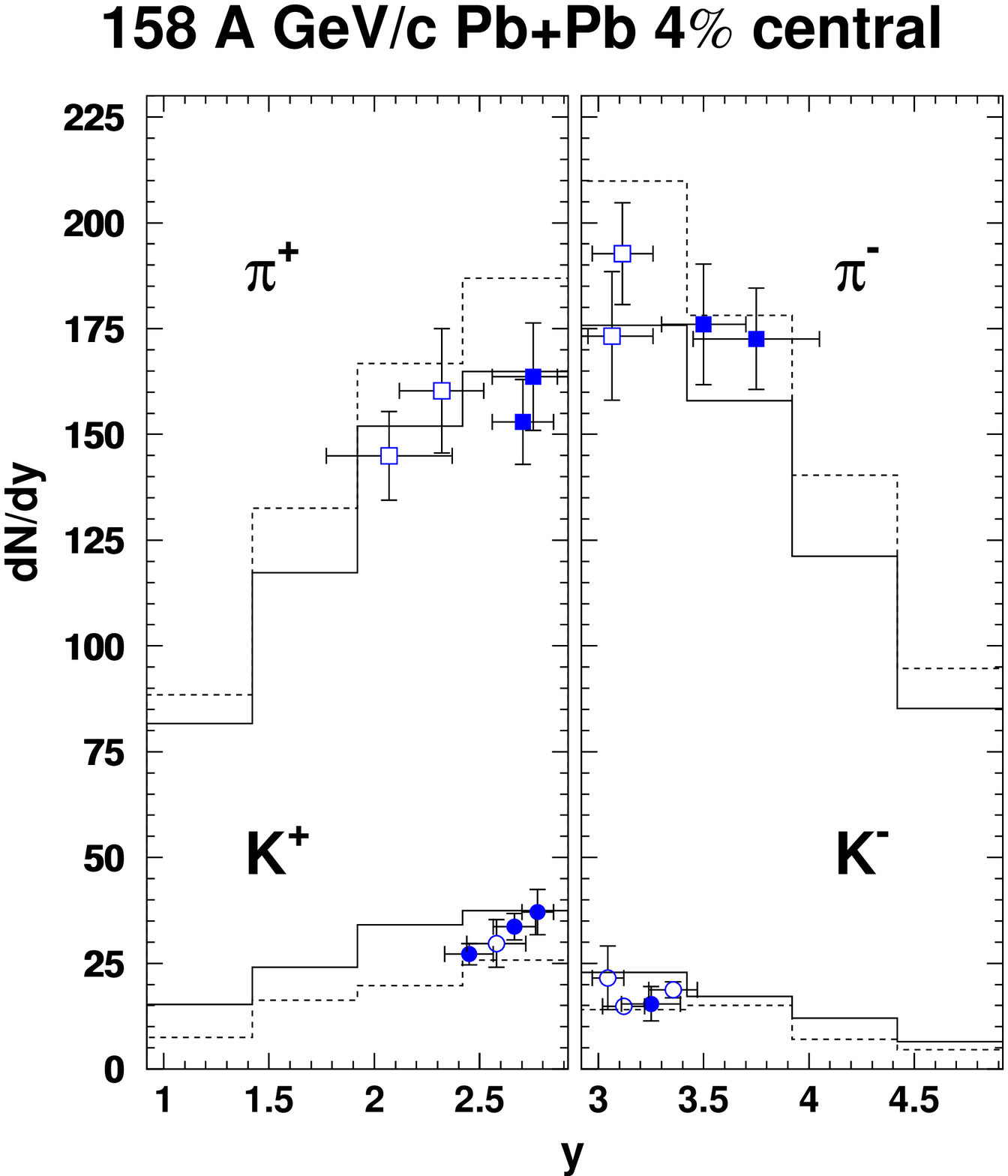}
\caption{
Comparison of measured charged kaon and pion yields with RQMD
 predictions.
The vertical error bars indicate statistical and systematic errors,
added in
quadrature; the horizontal ones -- $y$ boundaries of the 
acceptance used for $p_T$ integration in each spectrometer setting.
Open symbols
 represent spectrometer settings whose $y$ position is
shown mirror-reflected around midrapidity (2.92);
their solid analogs -- the actual settings.
RQMD: solid line -- standard mode, dashed line -- no rescattering.
}
\label{dndy_04}
\end{figure}

Tables \ref{sigma_T} and \ref{sigma_dndy} give the 
$m_T$ slope parameters 
and values of $dN/dy$ for kaons and pions, along with the statistical
and systematic uncertainties. 
The measured distributions for charged kaons 
of both signs in transverse kinetic energy and
rapidity, are shown on Fig. \ref{kt} and Fig. \ref{dndy_04}, 
respectively.
\begin{table}
\caption{Inverse slope parameters T. 
\label{sigma_T}}
\begin{tabular}{cccl}
\hline
 PID & $y$ interval &   $T$ (MeV) & $ \sigma(T)$ stat., syst. (MeV)\\
\hline
 $K^+$          & 2.3-2.6 & 230         & $\pm$ 8 $\pm$ 14 \\
  $K^+$         & 2.4-2.9 & 254        &  $\pm$ 4 $\pm$ 7 \\
  $K^-$         & 2.3-2.6 & 259        & $\pm$ 8 $\pm$ 12 \\
  $K^-$         & 2.4-2.9 & 245         &  $\pm$ 7 $\pm$ 6 \\
\hline  
\end{tabular}
\end{table}

\begin{table}
\caption{Particle distributions in rapidity. Every spectrometer
 setting
 provides an independent measurement. Settings overlapping in $y$ 
 are listed  separately.
\label{sigma_dndy}}
\begin{tabular}{cccc}
\hline
 PID & $y$ interval &    $dN/dy$  &   $\sigma(dN/dy)$ \\
\hline
4\% centr. $K^+$  & 2.7-2.9   &  37.1      & $\pm$  5.4  \\
                  & 2.3-2.6   &  27.2      &  $\pm$  2.5     \\
                  & 3.1-3.4   &  29.7     &   $\pm$  5.6    \\
                  & 2.6-2.8   &  33.6      &  $\pm$   3.1    \\
4\% centr. $K^-$  & 2.7-2.9   &  21.5      &   $\pm$  7.5    \\
                  & 2.3-2.5   &  18.7      &   $\pm$  1.9    \\
                  & 3.1-3.4   &  15.4      &   $\pm$  4.1    \\
                  & 2.6-2.8   &  14.8      &   $\pm$  1.4    \\

4\% centr. $\pi^+$  & 3.3-3.7  &  160      & $\pm$   15   \\
 
                    & 2.6-2.9  &  153      & $\pm$   10   \\
                    & 3.5-4.0  &  145      & $\pm$   10  \\
                    & 2.6-2.9  &  164      & $\pm$   13   \\
4\% centr. $\pi^-$ & 3.3-3.7  &   176      & $\pm$   14  \\
                   & 2.6-2.9  &   193	   & $\pm$   12  \\
                   & 3.5-4.0  &   173      & $\pm$   12  \\
                   & 2.6-2.9  &   173      & $\pm$   15  \\

\hline
\end{tabular}
\end{table}
The $1/m_T$  scaled spectra 
look approximately exponential in accordance with the 
behaviour typical for thermalized ensembles of interacting particles,
or for particles in whose production the phase-space constraints 
played the dominant role \cite{mil-roz}.
The spectra were fit with an exponential in $(m_T - m)$, and the
 resulting
slopes are shown in the inserts in Fig. 1.
The inverse slopes of the $K^+$ and $K^-$ spectra are the same, 
within errors. 
Our event selection is sufficiently central that the slopes
show no dependence on multiplicity.

In Fig. 2, it is clear that many fewer kaons are produced than pions,
as was observed in $p+p$ collisions. There are approximately twice
as many positive as negative kaons produced. This is typical for
baryon rich systems, and was also observed in $p+p$ collisions. 
Preliminary NA49 measurements of $K^+$ and $K^-$ $\,dN/\,dy$ 
\cite{NA49s} are consistent with those reported here.

Both Fig. 1 and 2 compare the data with predictions of the  
transport theoretical approach RQMD \cite{RQMD}.
While RQMD tends to overpredict
both the $K^+$ and $K^-$ yields, 
for $K^-$ the discrepancy appears to be larger.
Running RQMD in the mode which does not allow the hadrons to rescatter 
(shown by the dashed line on the figure) decreases
the number of kaons produced. This result illustrates the importance
of the secondary scattering to the total kaon yields.
 Measurements of proton production at midrapidity\cite{NA44p}
and of the $p-\bar{p}$ rapidity distribution\cite{NA49p}
 indicate that RQMD somewhat overpredicts the degree of baryon 
 stopping. 
Because $\pi N$ inelastic collisions can produce kaons,
an increase in stopping translates naturally into kaon enhancement
 at midrapidity.
The data show that the hadron chemistry via secondary scattering, as
implemented in RQMD, successfully reproduces the general trends in the
hadron distribution. However, the hadron chemistry in the model
is not quantitatively correct. 

Exothermic strangeness exchange reactions of the kind
$K+N \rightarrow Y + \pi$ and
$\overline{K}+\overline{N} \rightarrow \overline{Y} + \pi$  
are favoured by the cooling of the system, and
redistribute strangeness from the mesonic to the  baryonic sector.
If larger systems reach lower temperature before freezing out 
\cite{Shuryak},
such reactions may be important in Pb+Pb collisions.
These strangeness processes are in RQMD.
The model seems to underpredict (preliminary)
$\Lambda$ yields \cite{NA49s} and overpredict $K^-$.
This may indicate that the details of the description should be
 reexamined.

\begin{figure}
\epsfxsize=7.5cm 
\epsfbox{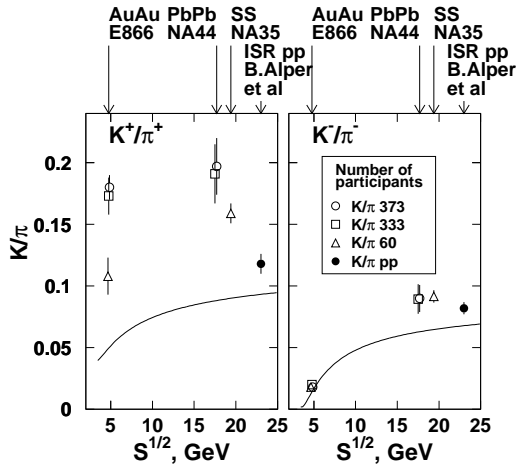}
\caption{
$K/\pi$ ratios in symmetric systems at midrapidity
 $|y-y_{CM}|\le |y_{proj}-y_{targ}|/8$.
The solid line shows full solid angle $K/\pi$ in $p+p$ collisions 
from the interpolation \protect{\cite{Rossi}}.
The data points from other experiments result from an interpolation
in $y$ to the midrapidity interval.
The E866 data points \protect{\cite{E802}} are also interpolated 
in the number of participants, for comparison with the SPS data.
}
\label{kpi_s}
\end{figure}

\begin{figure}
\epsfxsize=7.5cm 
\epsfbox{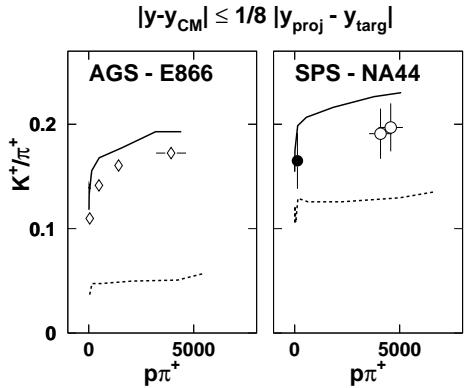}
\caption{Comparison of measurements with RQMD predictions:
 $K^+/\pi^+$ ratio in the specified rapidity
interval around mid-rapidity, as a function of the product
of pion and proton $dN/dy$, 
obtained in the same rapidity interval,  in symmetric
collisions. 
\large $\diamond$ \normalsize -- E866 AuAu, 
\large $\bullet$ \normalsize -- NA44 SS, 
\large $\circ$  \normalsize -- NA44 PbPb.
RQMD: solid line -- standard mode, dashed line -- no rescattering.
}
\label{p_pi}
\end{figure}

The $K/\pi$ abundance ratio allows estimation of  the degree of
 strangeness
enhancement and comparison of various colliding systems.
Fig. \ref{kpi_s} summarizes the existing midrapidity data in
 symmetric systems: ISR p+p \cite{Alper}; 
AGS AuAu \cite{E802}; SPS SS \cite{NA35SS}; and SPS PbPb.
\footnote{In this figure, as well as in Fig. \ref{p_pi}, the 
hadron abundances we present are integrals over a fixed fraction
 of rapidity around
midrapidity $y_{CM}$: $|y-y_{CM}|\le |y_{proj}-y_{targ}|/8$.
This enables the comparison between various energies and experiments,
but involves an interpolation in $y$ for 
experiments with larger coverage.}
Strangeness enhancement compared to the interpolated
\cite{Rossi} $pp$ collision data, shown as the line, is seen.
The solid point, corresponding to ISR data at midrapidity, indicates 
the extent of the enhancement due to the midrapidity cut on the 
particles.
The figure shows that $K^+/\pi^+$ is enhanced in high multiplicity
heavy ion collisions, but $K^-/\pi^-$ is consistent with  $p+p$
values. Higher multiplicity, or more central collisions, yields larger
enhancement, independent of $\sqrt{s}$.

Secondary hadronic interactions 
of the type  $\pi + N \rightarrow Y + \overline{K}$
are important for
the strangeness production \cite{RQMDf,Sorge},
and their rate is proportional
to the product of the participant's effective concentrations.
Fig. \ref{p_pi} shows 
the dependence of the $K^+/\pi^+$ ratio on the product of rapidity
densities of the two ingredients of the associated strangeness
production, $N$ (represented by $p$)
and $\pi^+$ in the AGS \cite{E866} and SPS \cite{NA44SS} data,
and RQMD calculations.
This ``$p\times\pi$'' product serves as an
observable measure of the strangeness-enhancing rescattering. 
The rate of change in the  $K^+/\pi^+$ ratio
with this rescattering observable
is initially very high. However, $K^+/\pi^+$ nearly saturates after
this initial rise. The figure shows why the enhancement is large as
soon as the multiplicity becomes appreciable. The values of 
``$p\times\pi$'' reached at the SPS and AGS are comparable, explaining
the similarity of the kaon enhancement despite the different energies.
RQMD reproduces the trend of the data very well, and the dotted
lines (illustrating no rescattering) along with the shape of the rise 
with ``$p\times\pi$'' underscore
the role of hadronic rescattering in kaon yields. The quantitative
agreement of RQMD with the data is not as good, but the final results
are undoubtedly quite sensitive to the magnitude of the cross sections 
used in the model.

\section{Conclusions} 
Production of charged $K$ and $\pi$ mesons in central Pb+Pb collisions
at 158 GeV/nucleon has been measured. Within the centrality range 
studied,
no strong multiplicity dependence of the kaon $m_T$ slopes or
$K/\pi$ ratios has been observed. We see no significant slope
 difference
between $K^+$ and $K^-$. $K^+/\pi^+$ is enhanced by a factor of about 
two over $p+p$ collisions, whereas $K^-/\pi^-$ is little enhanced. Our
measurement of $K^+/K^-$ in this saturated region may be used for
chemical calculations of the hadron gas.

Comparison with the RQMD model shows that the model qualitatively
reproduces the hadron chemistry, through the rescattering of the
produced particles. 
Quantitative comparisons, however, show that the model
overpredicts the $K^-$, while the magnitude of $K^+$ enhancement 
is within the range explainable by the RQMD mechanisms.
Deconfinement scenarios of the $K^+/\pi^+$ enhancement can not,
however, be ruled out or proven by these data alone.

\section{Acknowledgements}
We are grateful to Heinz Sorge for many helpful and illuminating 
conversations.
The NA44 Collaboration wishes to thank the staff of the CERN PS-SPS 
accelerator complex for their excellent work, and the technical
staff in
the collaborating institutes for their valuable contributions.
This work was supported by the Science Research Council of Denmark;
the Japanese Society for the Promotion of Science; the Ministry of
Education, Science and Culture, Japan;  the Science Research Council 
of Sweden; the US Department of Energy 
and the National Science Foundation.


\begin{thebibliography}{99}

\bibitem{RQMDf}
H. Sorge, Phys.\ Rev.\ C {\bf 52} (1995) 3291 
and references to other models therein.

\bibitem{QGP}
J. C. Collins, M. J. Perry, Phys. Rev. Lett. {\bf 34} (1975) 1353.

\bibitem{RafHag}
 J. Rafelski,  R. Hagedorn, in: Statistical Mechanics of
Quarks and Hadrons (Bielefeld, August 1980), 
ed.  H.Satz (North Holland, Amsterdam, 1981) p 253;
 J. Rafelski, B. M\"uller, Phys.\ Rev.\ Lett. {\bf 48} (1982) 1066.

\bibitem{bel}
S. Z. Belenky, Nucl.\ Phys.\ {\bf 2} (1956) 259.

\bibitem{koch}
P. Koch and J. Rafelski, Nucl.\ Phys.\ {\bf A444} (1985) 678-691.

\bibitem{NA44ex}
H. Beker {\it et al.}, (NA44 Collaboration),
Phys.\ Lett.\ {\bf B302} (1993) 510.

\bibitem{bc}
N. Maeda {\it et al.}, NIM {\bf A346} (1994) 132-136.

\bibitem{UCAL}
T. Akesson {\it et al.}, NIM {\bf A241} (1985) 17-42.

\bibitem{RQMD} 
our comparison relies 
on the version 2.4 of the RQMD model.

\bibitem{NA49h-}
P. G. Jones and the NA49 Collaboration, 
Nucl.\ Phys.\ {\bf A610} (1996) 188c-199c. 

\bibitem{mil-roz}
G. A. Milekhin, I. L. Rozental, 
Sov.\ Phys.\ JETP {\bf 6} No 1 (1958) 154.

\bibitem{NA49s}
C. Bormann for the NA49 Collaboration, 
J.\ Phys.\ G:\ Nucl.\ Part.\ Phys. 
{\bf 23} (1997)  1817-1825.


\bibitem{NA44p}
I. G. Bearden {\it et al.}, Phys.\ Lett.\ {\bf B388} (1996) 431.

\bibitem{NA49p}
NA49 Collaboration, Phys.\ Rev.\ Lett.\ {\bf 82} (1999) 2471-2475.

\bibitem{Shuryak}
E. V. Shuryak, Phys.\ Lett.\ {\bf B207} No 3 (1988) 345-348.


\bibitem{Alper}
B. Alper et al, Nucl. Phys. {\bf B100} (1975) 237-290.  

\bibitem{E802}
F. Wang for the E-802 Collaboration, HIPAGS-96 (WSU-NP-96-16);
F. Wang, Ph. D. thesis

\bibitem{NA35SS}
The NA-35 Collaboration, Z.\ Phys.\ {\bf C58}, 367-375 (1993)

\bibitem{Rossi}
Rossi {\em et al.}, Nucl.\ Phys.\ {\bf B84} (1975) 269.

\bibitem{Sorge}
H. Sorge, Nucl.\ Phys.\ {\bf A630} (1998) 522c-534c. 

\bibitem{E866}
Y. Akiba {\em et al.} Nucl.\ Phys.\ {\bf A610} (1996) 139c-152c.  

\bibitem{NA44SS}
H. B{\o}ggild {\em et al.}, Phys.\ Rev.\ {\bf C59} (1999) 328-335. 

\end{thebibliography}
\end{document}